\documentclass{iaus}
\usepackage{graphicx}

\title[Submillimeter Masers] 
{Submillimeter \& Millimeter Masers}

\author[Humphreys]   
{E. M. L. Humphreys}

\affiliation{Harvard-Smithsonian CfA, 60 Garden Street, 
Cambridge, MA 02138, USA}

\pubyear{2007}
\volume{242}  
\pagerange{119--126}
\date{?? and in revised form ??}
\jname{Proceedings Title IAU Symposium}
\editors{A.C. Editor, B.D. Editor \& C.E. Editor, eds.}

\begin{document}

\maketitle

\begin{abstract}
Despite theoretical predictions of the existence of many submillimeter masers, 
and some pioneering observational discoveries over the past few decades, 
these lines have remained relatively unstudied due to (i) challenges 
associated with observing at shorter wavelength; and, (ii) lack of 
possibility of high ($<$ 14$''$ at 345~GHz) angular resolution 
observations. With the advent of the {\sc SMA}, the first submillimeter imaging array capable of 
sub-arcsecond resolution, {\sc APEX}, and the promise of {\sc ALMA}, 
opportunities are opening for performing new science with 
millimeter/submillmeter masers. In this talk, I will review recent 
work in the field - including extragalactic H$_{2}$O millimeter masers, 
hydrogen recombination masers, submillimeter masers in star-forming regions,
and in the envelopes of evolved stars - and discuss prospects for the future.  
\end{abstract}

\section{Introduction}

Submillimeter masers exist in a wide range of astronomical environments, and provide the
possibility to probe physical conditions, source dynamics and magnetic fields on small angular scales.
They occur in several molecular and atomic species, including H$_{2}$O, SiO, H (recombination), 
CH$_{3}$OH, HCN, and SiS, and can be very strong (e.g., 8000~Jy for the 325~GHz H$_{2}$O masers in W49N; \cite{Menten1990a}). 
However, lack of angular resolution at submillimeter wavelengths has, until recently,
been a serious obstacle to realizing the potential of the masers.
Relating cm-wave maser emission observed on, say  0$\rlap{.}''$001  scales,    
with that of the submillimeter maser emission on $>$10$''$ scales (at 345~GHz), has made it difficult 
to constrain and test the radiative
transfer models that we will need to use in the Atacama Large Millimeter Array ({\sc ALMA}) era to map out precise
source temperature and density distributions.

The Submillimeter Array ({\sc SMA}) on Mauna Kea, operating from 0.3 to 2~mm,
is the first instrument capable of imaging in the submillimeter on sub-arcsecond scales 
(0$\rlap{.}''$25  at 345~GHz), 
and {\sc ALMA} will further transform maser science opportunities (see review
by Wootten in these proceedings). 
In this review, I will discuss results for masers at wavelengths shorter than 1.6~mm
($\nu$ $>$ 180~GHz), and future prospects for their observation using e.g., {\sc ALMA}, the {\it Herschel} satellite, 
the Stratospheric Observatory for Infra-Red Astronomy ({\sc SOFIA})
and submillimeter Very Long Baseline Interferometry ({\sc VLBI}).

\section{(Sub)millimeter H$_{2}$O Masers}\label{sec:water}

The H$_{2}$O masers detected to date, from rotational transitions within
the vibrational ground state and within the $\nu_2$=1 bending mode, are listed 
in Table~1 and are marked on the energy level diagram in Figure~1. 
The most studied lines are those at 183, 321 and 325~GHz, despite the 
relatively low atmospheric transmission at 183 and 325~GHz due to their low energies above ground state (Figure~2).
These masers are believed to be collisionally-pumped by a subset of the conditions that pump 22~GHz
masers, for the parameter space investigated by \cite[Neufeld \& Melnick (1991, hereafter NM91)]{Neufeld1991} and by \cite[Yates, Field \& Gray (1997, hereafter YFG97)]{Yates1997}. However, 
\cite[Cernicharo \etal\ (1994, 1999, 2006a, 2006b)]{Cernicharo1994} find
that the 183 and 325~GHz transitions can also be inverted in significantly lower temperature and density regimes of T$_{k}$$\sim$ 40~K and n(H$_{2}$)=10$^{5}$-10$^{6}$ cm$^{-3}$. NM91 and YFG97 are in broad agreement, except that YFG97
find that the masers at 439 and 470~GHz are pumped by radiation
from warm dust.
Both NM91 and YFG97 make predictions for new H$_{2}$O masers (e.g., at 448, 1296, \& 1322~GHz; 
also \cite{Deguchi1977}, \cite{Cooke1985}, \cite{Deguchi1990}, \cite{Humphreys2001}),
some of which could be observed using {\it Herschel}.
NM91 and YFG97 do not include levels from the $\nu_2$=1 vibrationally-excited state, see    
\cite[Deguchi (1977)]{Deguchi1977}, \cite[Deguchi \& Nguyen-Q-Rieu (1990)]{Deguchi1990} and
\cite[Alcolea \& Menten (1993)]{Alcolea1993} for $\nu_2$=1 maser pumping models.
Modeling of the $\nu_2$=1 masers is severely hampered by lack of relevant collisional excitation rates.

\begin{table}\def~{\hphantom{0}}
  \begin{center}
  \caption{H$_{2}$O Masers}
  \label{tab:kd}
  \begin{tabular}{rcccccccl}\hline
 Freq.  & Transition                           & Vib.  & Species$^{1}$ &E$_{u}$/k &CSE$^{2}$ & SFR$^{2}$ & EXG$^{2}$& Primary Reference\\
 (GHz)  & J$_{k_a,k_c}$ - J$_{k_a,k_c}$  & State &         & (K)&    &     &    &       \\
\hline
 22.235 & 6$_{16}$ - 5$_{23}$                  & G     &   O     &644 & Y  &  Y  & Y  & \cite[Cheung \etal\ (1969)]{Cheung1969}\\
 96.261 & 4$_{40}$ - 5$_{33}$                  &$\nu_2$=1&   P     &3065 & Y  &     &    & \cite[Menten \& Melnick (1989)]{Menten1989}\\
\hline
183.308 & 3$_{13}$ - 2$_{20}$                  &G      &   P     &205 & Y  &  Y  & Y  &\cite[Waters \etal\ (1980)]{Waters1980}\\
232.687 & 5$_{50}$ - 6$_{43}$                  &$\nu_2$=1&   O     &3463 & Y  &     &    &\cite[Menten \& Melnick (1989)]{Menten1989} \\
293.439 & 6$_{61}$ - 7$_{52}$                 &$\nu_2$=1&   O     &3935 & Y  &     &    &\cite[Menten \etal\ (2006)]{Menten2006} \\
321.226 & 10$_{29}$ - 9$_{36}$                &G      &   O     &1862 & Y  &  Y  &    & \cite[Menten \etal\ (1990a)]{Menten1990a}  \\
325.153 &  5$_{15}$ - 4$_{22}$                 &G      &   P     &470 & Y  &  Y  &    & \cite[Menten \etal\ (1990b]{Menten1990b}  \\
$^{3}$336.228 &  5$_{23}$ -  6$_{16}$                &$\nu_2$=1&   O      &2956 & Y  &     &    &\cite[Feldman \etal\ (1993)]{Feldman1993}\\ 
354.885 & 17$_{4 12}$ - 16$_{7 10}$            & G     &   O     &5782 & Y  &     &    & \cite[Feldman \etal\ (1991)]{Feldman1991}\\
380.194 & 4$_{14}$ - 3$_{21}$                  & G     &  O     &324 &    &  Y  &    & \cite[Phillips \etal\ (1980)]{Phillips1980}\\
437.347 & 7$_{53}$ - 6$_{60}$                  &G      &  P       &1525 & Y  &    &    & \cite[Melnick \etal\ (1993)]{Melnick1993}\\
439.151 & 6$_{43}$ - 5$_{50}$                  &G      &  O       &1089 & Y  &  Y  &    & \cite[Melnick \etal\ (1993)]{Melnick1993}\\
470.889 & 6$_{42}$ - 5$_{51}$                  &G      &  P       &1091 & Y  &  Y  &    & \cite[Melnick \etal\ (1993)]{Melnick1993}\\
658.007 & 1$_{10}$ - 1$_{01}$                  &$\nu_2$=1&   O     &2361 & Y  &     &    &  \cite[Menten \& Young (1995)]{Menten1995}\\\hline
  \end{tabular}
 \end{center}
$^{1}$ O=ortho-H$_{2}$O (parallel hydrogen atom nuclear spins)
and P=para-H$_{2}$O (anti-parallel hydrogen nuclear spins). 
In thermal equilibrium, the two forms are present in an O/P ratio of 3:1.\\ 
$^{2}$ CSE=Circumstellar Envelope; SFR=Star Forming Region; EXG=Extragalactic\\
$^{3}$ Quasi-maser (\cite{Feldman1993}), or thermal  (\cite{Menten2006}), emission toward VY CMa.
\end{table}

\subsection{H$_{2}$O Masers in Evolved Stars}

(Sub)millimeter masers at 183, 321 and 325~GHz are common 
in the circumstellar envelopes (CSEs) of evolved stars. 
70\% of the 22~GHz H$_{2}$O maser sources observed by \cite[Yates, Cohen \& Hills (1995)]{Yates1995} 
also have H$_{2}$O maser emission at 321 and 325~GHz. 
Single-dish linewidths of 22 and 325~GHz masers have 
similar extents and peak flux densities, whereas 321~GHz maser
 \begin{figure}
\includegraphics[height=6.325in,width=4.8875in]{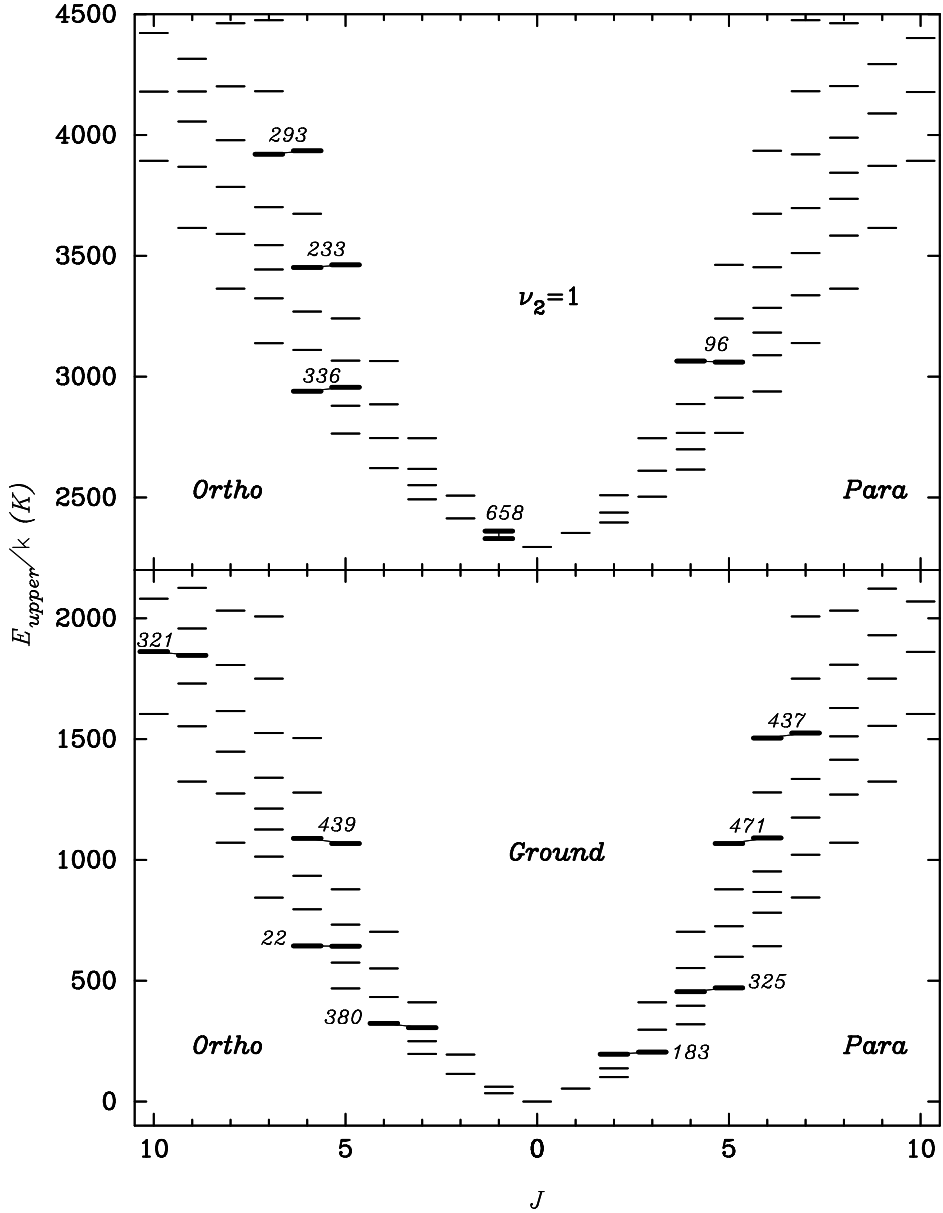}
\caption{H$_{2}$O Energy Level Diagram. Rotational levels in the ground
and vibrationally-excited $\nu_2$ states are shown for energies between 0 to 2200~K, and
2200 to 4500~K respectively. Levels of maser transitions are plotted in bold, 
and labels are the transition frequencies in gigahertz. The maser at 355~GHz, at an
energy of 5782~K above ground-state, is not shown on this plot.  
Ortho-H$_{2}$O plotted for values of the total molecular angular momentum $J$ increasing to the left, 
para-H$_{2}$O to the right. Data are from the experimentally-derived energy levels
of \cite[Tennyson \etal\ (2001)]{Tennyson2001}, available on http://www.tampa.phys.ucl.ac.uk/ftp/astrodata/water/levels.
 \label{fig:contour}}
\end{figure}
line widths are narrower and weaker by a factor of a few (an exception
is emission from R~Aqr, a Mira variable in a symbiotic binary; \cite{Ivison1998}).
321~GHz emission likely originates from a subset of the conditions that
give rise to the 22 and 325~GHz emission, close to the central star.
The 321~GHz line is generally more variable than the 22 and 325~GHz emission and 
variations in the 22, 321 and 325~GHz masers are not particularly well-correlated (in some cases they are completely anticorrelated;
 \cite{Yates1996}).
For 183~GHz H$_{2}$O masers,  \cite[Gonz\'{a}lez-Alfonso \etal\ (1998)]{Gonzalez-Alfonso1998}
find that variability of the line profile and flux from one epoch to another is small in 
comparison with that of 22~GHz masers in a study of 23 evolved stars. 
As for 22~GHz H$_{2}$O masers, in stars of low mass-loss rates ($\dot{M}$) the 
183~GHz emission peaks at a velocity similar to that of the star, 
whilst in stars with high $\dot{M}$ the emission peaks at velocities closer to the terminal 
velocity of the envelope (tangential vs. radial amplification as the envelope becomes denser at greater radii). Masers at 437, 439 and 471~GHz have all been detected in CSEs and the
437~GHz line has been found exclusively in this environment (\cite{Melnick1993}; YFG97). 
Masers from the $\nu_2$=1 state, at 96, 233, 293, (336) \& 658~GHz,  
are only known to occur strongly from CSEs, although 658~GHz emission of undetermined nature is observed 
toward Orion-KL (\cite{Schilke2001}). On the basis of excitation arguments, 
and similarity with SiO maser lineshapes in some cases, the $\nu_2$=1 masers
likely occur close (within a few R$_{*}$) to the central star.
In recent Atacama Pathfinder Experiment (APEX) observations towards VY CMa, \cite[Menten \etal\ (2006)]{Menten2006}
find weak maser emission from the 293~GHz line, a non-detection of emission at 297~GHz and thermal emission at 336~GHz.
For a discussion of SMA observations of the 658~GHz masers, often particularly strong e.g., 3000 Jy toward VY CMa, see the review by Hunter in these proceedings.

\begin{figure}
\hspace{0.25cm}
\includegraphics[height=5.225in,width=4.0375in,angle=-90]{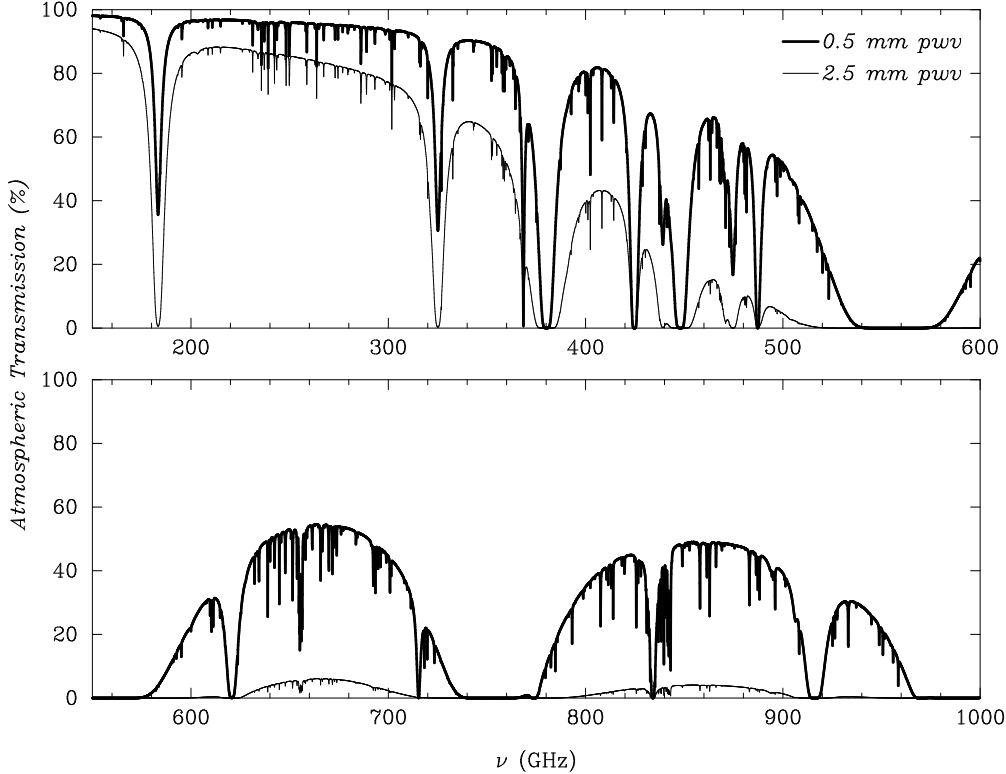}
\caption{Zenith atmospheric transmission at Mauna Kea for column densities of 0.5 and 2.5~mm 
H$_{2}$O ($\tau_{225GHz}$$\sim$0.04 and 0.13 respectively). Data are from the Caltech Submillimeter Observatory Atmospheric
Transmission Interactive Plotter (http://www.submm.caltech.edu/cso/weather/atplot.shtml).
}
\end{figure}

\subsection{H$_{2}$O Masers in Star-Forming Regions}

\begin{table}\def~{\hphantom{0}}
  \begin{center}
  \caption{Some (Sub)millimeter H$_{2}$O Observations Towards Star-forming Regions}
  \label{tab:kd}
  \begin{tabular}{llcl}\hline
Freq. & Sources   & Telescope$^{1}$          &Comments                          \\
(GHz) &           & (Beam)          &                                   \\\hline
{\bf 183}   & Orion-KL  & KAO  (7$\rlap{.}'$5)              &First 183 GHz detection (\cite{Waters1980})\\
      & Orion-KL, Cep A,      & IRAM &Established 183 GHz maser emission widespread \\
      & W49N, S252A, S158, &   30-m          & (\cite{Cernicharo1990})         \\
      & HH7-11A, W3(H2O) &  (14$''$)           &        \\
      &NGC~7538S, RN013      &             &            \\
      &Orion          &  IRAM       & Spatially-extended emission; strong, narrow \\
      &               &   30-m      &  features at IRC2 (\cite{Cernicharo1994})\\
      &W49N           &  IRAM       &Spatially-extended; less time-variable than at \\ 
      &               &  30-m       & 22 \& 325 GHz (\cite{Gonzalez-Alfonso1995}) \\ 
      &HH7-11A, L1448IRS3,& IRAM        & 183~GHz maser variability in low-mass  star           \\
      &    L1448-mm     &  30-m       & formation \cite[Cernicharo \etal\ (1996)]{Cernicharo1996}            \\ 
      & Sgr B2        &  IRAM           & Strong toward cores; moderate emission at Sgr \\
      &               &  30-m         &  B2 main condensations (\cite{Cernicharo2006a}) \\\hline
{\bf 321}   &W3(OH), W49N,    &  CSO        & Strongest 22 GHz \& 321 GHz features generally  \\   
      & W51 IRS2 \& Main         &   (23$''$)  & at similar velocities  (\cite{Menten1990a})   \\  
      &Cep A &    SMA   & 22 \& 321 GHz distributions perpendicular (cm  \\
      &          & (0$\rlap{.}''$75)    & \& submm obs. $\sim$1 mth apart) (\cite{Patel2007}) \\\hline
{\bf 325}   & Orion-KL &        CSO                    &22 \& 325 GHz cover similar velocity extents\\
      &  W49N, W51 Main         &      (22$''$)                      &(\cite{Menten1990b})\\
   & IRAS 16293-2422& & \\
   & G34.3-0.2, W49N,& CSO   & 325, 439 \& 470 GHz cover similar velocity extents    \\
   & Sgr B2 &      & (\cite{Melnick1993})    \\
   & Orion-KL & CSO    &  325 GHz emission much less extended than at  \\
   &          &        &  183~GHz  (\cite{Cernicharo1999})\\
   & Orion-KL        & SMA  & In high-mass protostar Source I outflow, 325~GHz   \\
   &                 & (0$\rlap{.}''$65) & emission    more collimated than 22~GHz (cm \& \\
&                    & Full Stokes&  submm obs. $\sim$5 yrs apart)  (Greenhill \etal\ 2007)\\\hline
{\bf 439,} & G34.3-0.2        &  CSO       &   First detections: 325, 439 \& 470 GHz cover         \\
{\bf 471}   & W49N, Sgr B2           &  (16$''$)       & similar velocity extents   (\cite{Melnick1993})                               \\\hline
  \end{tabular}
 \end{center}
$^{1}$ KAO = Kuiper Airborne Observatory; IRAM = Institut de Radioastronomie Millim\'{e}trique; 
CSO = Caltech Submillimeter Observatory; SMA = Submillimeter Array
\end{table}

Observations of the (sub)millimeter H$_{2}$O masers are summarized in Table~2.
183, 321, 325, 439 and 471~GHz masers have been observed towards
high-mass star-forming regions, the 183 and 325~GHz lines have also been
observed towards low-mass star-forming regions (HH7-11A, L1448IRS3,  L1448-mm at 183~GHz; 
IRAS16293-2422 at 325~GHz). The velocity range covered by the 321~GHz maser is typically
smaller than that observed at 22, 183, and 325~GHz. The 321~GHz emission is typically weakest
of these four lines, and the 22~GHz is the strongest.
 
First arcsecond resolution observations of H$_{2}$O masers towards 
a star-forming region were performed at 325~GHz towards Orion-KL 
by \cite[Greenhill \etal\ (2007)]{Greenhill2007}
using the compact configuration of the SMA, and followed up with a higher resolution (0$\rlap{.}''$65 circular), 
full polarization epoch.  In previous mapping of this region using the Caltech Submillimeter Observatory (CSO) 
with a 22$''$ beam, \cite[Cernicharo \etal\ (1999)]{Cernicharo1999} 
concluded that the 325~GHz emission traces extended, low-density material of n(H$_{2}$)$\sim$10$^{5-6}$ cm$^{-3}$. 
However, \cite[Greenhill \etal\ (2007)]{Greenhill2007} find that 
it also arises from compact high-density clumps, much as the 22~GHz transition, although in the outflow
of high-mass protostar Source I the 325~GHz emission appears more collimated.  
Line ratios of these H$_{2}$O transitions could therefore be valuable diagnostics for shocked material in 
protostellar outflows.

Using the SMA, \cite[Patel \etal\ (2007)]{Patel2007} imaged 321 GHz H$_{2}$O maser emission towards 
high-mass star-forming region Cepheus A with a resolution of 0$\rlap{.}''$75, in close time proximity 
to Very Large Array observations of the 22 GHz H$_{2}$O masers (43 days later).
The majority of 321~GHz maser spots did not appear to be associated with those at 22~GHz, 
and the position angles of the roughly linear structures traced by the masers appeared perpendicular, perhaps
tracing a jet and disk respectively.
\cite[Patel \etal\ (2007)]{Patel2007} interpret the 
submillimeter masers in Cepheus A to be tracing significantly hotter regions (600-2000 K) than the 
centimeter masers, see the contribution by Patel in these proceedings for further details.

\subsection{Extragalactic H$_{2}$O Masers}

There have been two recent detections of extragalactic H$_{2}$O masers at 183~GHz. 
\cite[Humphreys \etal\ (2005)]{Humphreys2005}  detected emission toward the 
well-known 22~GHz H$_{2}$O megamaser galaxy NGC~3079 using the SMA.
At a distance of 16 Mpc, NGC~3079 harbors an active galactic nucleus (AGN), and 
additionally has some starburst indicators. Spatially and kinematically the
183~GHz emission is associated with the AGN, with emission peaking at the
same position as that of 22~GHz emission imaged by \cite[Kondratko \etal\ (2005)]{Kondratko2005} using VLBI.
At 22 GHz, the emission  has a time-variable peak flux density in the range 3-12 Jy, 
whereas at 183 GHz, the  H$_{2}$O maser emission had a peak flux density of $\sim$0.5 Jy.
\cite[Humphreys \etal\ (2005)]{Humphreys2005} also make a tentative detection of the 439~GHz
maser using the JCMT.

\cite[Cernicharo, Pardo \& Weiss (2006)]{Cernicharo2006} detected a megamaser at 183.310 GHz in Arp 220 using the IRAM 30~m, with a line width of 
$\sim$350 km s$^{-1}$ and total luminosity of $\sim$2.5 $\times$ 10$^8$ K km s$^{-1}$ pc$^2$. This is very interesting since no emission at 22 GHz 
has been detected from Arp~220 (an OH megamaser source). 
This fact puts constraints on the physical conditions of the central region of Arp 220,
 which are further strengthened by observations of HCN and HNC $J=3-2$ and $J=1-0$, suggesting
densities of n(H$_{2}$)=10$^{5}$ cm$^{-3}$. \cite[Cernicharo, Pardo \& Weiss (2006)]{Cernicharo2006}
propose a scenario with $\sim$10$^6$ star-forming cores similar to those found in Sgr B2 in the central 
kiloparsec of Arp 220. 
 The 183 GHz line is therefore an additional tool to explore the physical conditions in starburst and AGN sources, with
the potential for high angular resolution observations using {\sc ALMA}. 

\section{(Sub)millimeter SiO Masers in Evolved Stars}

(Sub)millimeter $^{28}$SiO masers have been detected from the 
$J=$5-4 $\approx$ 215~GHz ($v=1$ \& 2, \cite{Clemens1983}; $v=3$ tentative 
detection from VX Sgr, \cite{Jewell1987}; $v=$ 3 \& 4 from  VY CMa, \cite{Cernicharo1993}),
 $J=6-5$ $\approx$ 258 GHz ($v=$1, \cite{Jewell1987}; $v=$2, VY CMa, \cite{Cernicharo1993}),
$J=7-6$ $\approx$ 301 GHz ($v=1$ \& 2, R Aqr, \cite{Gray1995}), $J=8-7$ $\approx$ 344 GHz 
($v=1$, VY CMa, and tentative $v=2$, \cite{Humphreys1997}; $v=2$, VY CMa, \cite{Gray1999}). 
The highly-rotationally excited masers are very rare from the $v=3$ \& 4 states (\cite{Pardo1998} and 
references therein) which lie at $>$5400~K above ground state. They are more common in the $v=$1 \& 2
(\cite{Jewell1987}; \cite{Cernicharo1993}; \cite{Humphreys1997};  \cite{Gray1999}) especially
ing $J=$5-4 emission, but weaker than their lower frequency counterparts in the same vibrational
states, and more time-variable. In a survey of 34 supergiant and long-period variable stars,
\cite[Gray, Humphreys \& Yates (1999)]{Gray1999} found that for Mira variables, emission
from the high-frequency transitions is absent or weak from optical phase range $\phi$ $\sim$0.4 -- 0.7 of
the stellar pulsation cycle. 

SiO maser emission at lower frequencies is well-known to display high degrees (tens of \%) of linear polariation
e.g.,    $v=1$ $J=1-0$  (43~GHz) maser components can be $\sim$100\% linearly polarized (e.g., \cite{Kemball1997}). 
Using a partially-completed SMA, \cite[Shinnaga \etal\ (2004)]{Shinnaga2004} imaged the $v=1$, $J=5-4$ SiO maser emission of supergiant VY CMa 
to investigate linear polarization properties at higher frequency. The majority of components showed significant degrees of 
linear polarization, with one at the 60\% level, that Shinnaga et al. attribute to a radiative pumping process.

For the less abundant isotopomers $^{29}$SiO and $^{30}$SiO, 
\cite[Cernicharo \& Bujarrabal (1992)]{Cernicharo1992} detected
maser emission from the $v=0$ $J=5-4$ transition for both species,
 the $^{29}$SiO $v=2$, $J=6-5$ line, and the $^{30}$SiO $v=1$ $J=6-5$
towards VY CMa.
For $^{29}$SiO, the $v=3$ $J=8-7$ at 335.9~GHz was detected toward TX Cam, R Leo and W Hya
at optical stellar phases  $\phi$ of 0.3, 0.15 and 0.25 respectively (\cite{Gonzalez-Alfonso1996}) and 
towards  VY CMa (\cite{Gonzalez-Alfonso1996}; \cite{Menten2006} using APEX).
\cite[Menten \etal\ (2006)]{Menten2006} also detected maser emission in the $^{30}$SiO $v=1$ $J=8-7$
line towards VY CMa, whereas the $^{29}$SiO $v=0$ $J=8-7$ transition appears thermal.
Infra-red line overlaps of the SiO isotopomers is believed to be important
in the pump scheme of these masers (e.g., \cite{Herpin2000}). For a detailed discussion of SiO masers in evolved stars, see the review by Bujarrabal in these proceedings.

\section{(Sub)millimeter H Recombination Masers}

Hydrogen recombination maser emission is known from two galactic peculiar stellar sources, 
 MWC 349A (\cite{Martin-Pintado1989})  and Eta Carinae (\cite{Cox1995}).
(Sub)-millimeter maser emission from MWC~349A has been detected from at least the 
H31$\alpha$ (210.5 GHz), H30$\alpha$ (231.9 GHz),  H29$\alpha$ (256.302 GHz) (\cite{Martin-Pintado1989}) 
from  H26$\alpha$ (353.623 GHz; \cite{Thum1994a}) and the H21$\alpha$ (662.405 GHz; 350~Jy; \cite{Thum1994b}), 
H32$\beta$ (366.6 GHz; \cite{Thum1995}). \cite[Planesas, Martin-Pintado \& Serabyn (1992)]{Planesas1992} 
spatially resolved the double-peaked maser spectrum into two emitting regions, separated by 0$\rlap{.}''$065, associated
with the red and blue-shifted emission from a sub-arcsecond disk imaged in the near-infrared by \cite[Danchi, Tuthill \& Monnier (2001)]{Danchi2001}.  
\cite[Weintroub \etal\ (2007)]{Weintroub2007} again detected 
H30$\alpha$ and H26$\alpha$ maser emission from the two regiond using the SMA, but also found
emission at positions between them with  
an accuracy of 0$\rlap{.}''$01. The emission position-velocity diagram is consistent with that of an
edge-on disk in approximate Keplerian rotation. However, \cite[Weintroub \etal\ (2007)]{Weintroub2007} 
argue that systematic deviation from Keplerian rotation may indicate the 
presence of spiral structure in the MWC 349A disk (see also these proceedings).
From Zeeman observations of the H30$\alpha$ maser, \cite[Thum \& Morris (1999)]{Thum1999} 
report a dynamically-important magnetic field associated with the corona of the circumstellar disk, 
possibly generated by a local disk dynamo. 
Pumping of the masers in MWC 349A has been explained by \cite[Strelnitski \etal\ (1996)]{Strelnitski1996}.
Towards Eta Carinae, \cite[Cox \etal\ (1995)]{Cox1995} detected millimeter maser 
emission at H30$\alpha$, H29$\alpha$ and H37$\beta$ (240.021 GHz) (see also 
\cite[Abraham \etal\ (2002)]{Abraham2002}).

Extragalactic H recombination maser emission from the H27$\alpha$ (316.416 GHz) transition has also been 
detected towards M82 (\cite{Seaquist1996}). The emission is highly time-variable, 
and of peak flux density 1.5~Jy at the strongest epoch. 
We note that H recombination masers at lower frequency may also have been detected 
from starburst galaxies, see references in \cite[Seaquist \etal\ (1996)]{Seaquist1996}, 
and that H recombination masers are predicted to  probe the Epochs of Recombination and Reionization
(\cite{Spaans1997}).

\section{(Sub)millimeter CH$_{3}$OH Masers}

In a survey of Galactic star-forming regions, \cite[Kalenskii, Slysh \& Val'Tts (2002)]{Kalenskii2002}
detected maser emission from methanol 8$_{-1}$ - 7$_0$ E at 229.8 GHz towards DR 21(OH) and DR 21 West, and 
toward two maser candidates, L 379IRS3 and NGC 6334I(N). The maser emission in DR21(OH) and DR 21 West indicates gas 
kinetic temperatures of T$_{k}$ $\sim$ 50~K and densities of n(H$_{2}$) = 3 $\times$ 10$^{4}$ cm$^{-3}$. Towards 16 other sources, 
the emission detected from this line was thermal in nature.  
\cite[Sobolev \etal\ (2002)]{Sobolev2002} reported the detection of class II methanol emission at 216.9 GHz, and 
models by \cite[Cragg \etal\ (2005)]{Cragg2005} predict the existence of many more (sub)millimeter Class II
methanol masers.  

\section{(Sub)millimeter HCN \& SiS Masers in Carbon Stars}

(Sub)millimeter HCN maser emission has been detected from carbon-rich circumstellar envelopes.
Using the Caltech Submillimeter Observatory (CSO), \cite[Schilke, Mehringer \& Menten (2000)]{Schilke2000} and 
\cite[Schilke \& Menten (2003)]{Schilke2003} detected the $J=9-8$ maser of the 
(04$^{0}$0) vibrationally-excited state of HCN at a frequency of $\approx$804.751 GHz towards 
IRC+10216 (at two epochs of peak flux densities 1420 \& 840 Jy) and CIT 6 (110 Jy). 
The lower level of the maser is at 4200 K above ground state, such that emission should originate
from the innermost region of the CSEs ($<$ 3.5 R$_{*}$).  \cite[Schilke \& Menten (2003)]{Schilke2003}
also detected the (11$^{1}$0)-(04$^{0}$0), $J=10-9$ maser at 890.761 GHz towards IRC+10216
(at four epochs with peak flux densities of 6120, 4430, 9230, 900 Jy), CIT 6 (1090 \& 1150 Jy) and Y CVn (140 Jy).
In surveys using the Heinrich-Hertz-Submillimeter Telescope, \cite[Bieging, Shaked \& Gensheimer (2000)]{Bieging2000} and 
\cite[Bieging (2001)]{Bieging2001} discovered maser emission in the $J=3-2$ (265.886 GHz) and $4-3$ (354.505 GHz) transitions of the HCN 
$(01^{1c}0)$ vibrational bending mode toward five stars: R Scl, V384 Per, R Lep, Y CVn, and V Cyg (out of 12 observed).
Submillimeter HCN masers at 964 and 968 GHz are also predicted by \cite[Schilke \& Menten (2003)]{Schilke2003}, and could be detected using {\sc SOFIA}.

SiS masers were first discovered by \cite[Henkel, Matthews \& Morris (1983)]{Henkel1983} 
from the $v=0$, $J=1-0$ transition at 18 GHz toward carbon-rich star 
IRC+10216. 
(Sub)millimeter SiS maser emission was also detected toward IRC+10216 from the $v=0$, $J=11-10$ (199.672~GHz), 
$J=14-13$ (254.103~GHz) and $J=15-14$ (272.243~GHz) transitions by 
\cite[Fonfr\'{\i}a Exp\'{o}sito \etal\ (2006)]{Fonfria2006} using the IRAM 30-m. Line overlap is believed
to be important in the pumping scheme of the highly-rotationally
excited masers and they are thought to occupy 
$\sim$5 - 7 R$_{*}$ in the CSE of IRC+10216. 
Future high-resolution observations of the HCN and SiS masers using {\sc ALMA} 
will therefore yield new information on the dust formation zone of carbon stars.

\section{Summary \& Future Propects}

Observations of submillimeter masers at high angular resolution provide 
new means of studying stellar evolution, star formation and AGN/starburst
activity. Where different maser transitions trace the same gas, 
we will be able to place new constraints on radiative transfer models to determine
small-scale source temperature and density distributions. 
Where maser lines trace different regions of sources, we will be able to map
out more of source structures and dynamics than ever before. Submillimeter masers could be
particularly important probes of regions in which longer wavelength maser emission 
is subject to obscuration  e.g., due to free-free or synchrotron opacity.

The spatial resolution and sensitivity of {\sc ALMA} will revolutionize submillimeter science. 
There have also been huge strides in submillimeter {\sc VLBI}, with fringes obtained at 129, 147, and 230~GHz
(see e.g., \cite{Krichbaum2007}) and with imaging of SiO $J=3-2$ masers at 129 GHz in VY CMa
and several AGB stars already achieved (\cite{Doeleman2005}; Doeleman, private communication).
Within the next decade, observations of submillimeter masers are likely to become very much more commonplace
and, in conjunction with detailed modeling,  will yield a wealth of new and exciting avenues of research.

\begin{acknowledgments}
EH thanks Lincoln Greenhill, Preethi Pratap, Andrej Sobolev, Vladimir Strelnitski 
and Jonathan Weintroub for providing unpublished results, and Jim Moran for
helpful comments on this manuscript.
\end{acknowledgments}



\begin{thebibliography}{}

\bibitem[Abraham \etal\ (2002)]{2002}
{Abraham, Z., Damineli, A., Durouchoux, P., Nyman, L., McAuliffe, F.} 2002,
in: V. Migenes \& M.~J. Reid  (eds.),
\textit{Cosmic Masers: From Proto-Stars to Black Holes},
IAU Symposium, vol.\ 206, p.\ 234 


\bibitem[Alcolea \& Menten 1993]{Alcolea1993}
{Alcolea, J. \& Menten, K.~M.} 1993, 
in: A.~W. Clegg \& G.~E. Nedoluha  (eds.),
\textit{Astrophysical Masers},
Lecture Notes in Physics (Berlin: Springer Verlag), vol.\ 412, p.\ 399 

\bibitem[Bieging, Shaked \& Gensheimer 2000]{Bieging2000}
{Bieging, J.~H., Shaked, S., Gensheimer, P.~D.} 2000, \textit{ApJ}, 543, 897

\bibitem[Bieging 2001]{Bieging2001}
{Bieging, J.~H.} 2001, \textit{ApJ}, 549, L125

\bibitem[Cernicharo \etal\ 1990]{Cernicharo1990}
{Cernicharo, J., Thum, C., Hein, H., John, D., Garcia, P., \& Mattioco, F.} 1990, \textit{A\&A}, 231, L15


\bibitem[Cernicharo \& Bujarrabal 1992]{Cernicharo1992}
{Cernicharo, J. \& Bujarrabal, V.} 1992, \textit{ApJ}, 401, L109

\bibitem[Cernicharo, Bujarrabal \& Santaren 1993]{Cernicharo1993}
{Cernicharo, J., Bujarrabal, V., Santaren, J.~L.} 1993, \textit{ApJ} 407, L33

\bibitem[Cernicharo \etal\ 1994]{Cernicharo1994}
{Cernicharo, J., Gonz\'{a}lez-Alfonso, E., Alcolea, J., Bachiller, R., John, D.} 1994, \textit{ApJ}, 432, L59

\bibitem[Cernicharo, Bachiller \& Gonz\'{a}lez-Alfonso 1996]{Cernicharo1996}
{Cernicharo, J., Bachiller, R., Gonz\'{a}lez-Alfonso, E.} 1996, \textit{A\&A}, 305, L5

\bibitem[Cernicharo \etal\ 1999]{Cernicharo1999}
{Cernicharo, J., Pardo, J.~R., Gonz\'{a}lez-Alfonso, E.,
	Serabyn, E., Phillips, T.~G., Benford, D.~J.,
	Mehringer, D.} 1999, \textit{ApJ}, 520, L131

\bibitem[Cernicharo \etal\ 2006a]{Cernicharo2006a}
{Cernicharo, J., Goicoechea, J.~R., Pardo, J.~R.,  
	Asensio-Ramos, A.} 2006a, \textit{ApJ}, 642, 940

\bibitem[Cernicharo, Pardo \& Weiss 2006b]{Cernicharo2006b}
{Cernicharo, J., Pardo, J.~R., \& Weiss, A.} 2006b, \textit{ApJ}, 646, L49

\bibitem[Cheung \etal\ 1969]{Cheung1969}
{Cheung, A.~C., Rank, D.~M., Townes, C.~H., Thornton, D.~D., Welch, W.~J.} 1969, \textit{Nature}, 221, 626

\bibitem[Clemens \& Lane 1983]{Clemens1983}
{Clemens, D.~P. \& Lane, A.~P.} 1983, \textit{ApJ}, 266, L117

\bibitem[Cooke \& Elitzur 1985]{Cooke1985}
{Cooke, B. \& Elitzur, M.} 1985, \textit{ApJ}, 295, 175

\bibitem[Cox \etal\ 1995]{Cox1995}
{Cox, P., Martin-Pintado, J., Bachiller, R., Bronfman, L., 
Cernicharo, J., Nyman, L.-A., Roelfsema, P.~R.} 1995, \textit{A\&A}, 295, L39

\bibitem[Cragg \etal\ 2005]{Cragg2005}
{Cragg, D.~M., Sobolev, A.~M., Godfrey, P.~D.} 2005, \textit{MNRAS}, 360, 533

\bibitem[Danchi, Tuthill \& Monnier 2001]{Danchi2001}
{Danchi, W.~C., Tuthill, P.~G., \& Monnier, J.~D.} 2001, \textit{ApJ}, 562, 440

\bibitem[Deguchi 1977]{Deguchi1977}
{Deguchi, S.} 1977, \textit{PASJ}, 29, 669

\bibitem[Deguchi \& Nguyen-Q-Rieu 1990]{Deguchi1990}
{Deguchi, S., Nguyen-Q-Rieu} 1990, \textit{ApJ}, 360, L27


\bibitem[Doeleman \etal\ 2005]{Doeleman2005}
{Doeleman, S.~S., Phillips, R.~B., Rogers, A.~E.~E.,  \etal\ } 2005,
in: J. Romney \&  M. Reid (eds.),  \textit{Future Directions in High Resolution Astronomy}, 
Astronomical Society of the Pacific Conference Series, vol.\ 340, 
p.\ 605

\bibitem[Feldman \etal\ 1991]{Feldman1991}
{Feldman, P.~A., Matthews, H.~E., Cunningham, C.~T., Hayward, R.~H., Wade, J.~D., Amano, T., Scappini, F.} 1991, 
\textit{JRASC}, 85, 191

\bibitem[Feldman \etal\ 1993]{Feldman1993}
{Feldman, P.~A., Matthews, H.~E., Amano, T., Scappini, F., Lees, R.~M.} 1993,
in: A.~W. Clegg \& G.~E. Nedoluha  (eds.),
\textit{Astrophysical Masers},
Lecture Notes in Physics (Berlin: Springer Verlag), vol.\ 412, p.\ 65
 
\bibitem[Fonfr\'{\i}a Exp\'{o}sito \etal\ 2006]{Fonfria2006}
{Fonfr\'{\i}a Exp\'{o}sito, J.~P., Agundez, M., Tercero, B., Pardo, J.~R., Cernicharo, J.} 2006, \textit{ApJ}, 646, L127

\bibitem[Gonz\'{a}lez-Alfonso \etal\ 1995]{Gonzalez-Alfonso1995}
{Gonz\'{a}lez-Alfonso, E., Cernicharo, J., Bachiller, R., Fuente, A.} 1995, \textit{A\&A}, 293, L9

\bibitem[Gonz\'{a}lez-Alfonso \etal\ 1996]{Gonzalez-Alfonso1996}
{Gonz\'{a}lez-Alfonso, E., Alcolea, J., Cernicharo, J.} 1996, \textit{A\&A}, 313, L13

\bibitem[Gonz\'{a}lez-Alfonso \etal\ 1998]{Gonzalez-Alfonso1998}
{Gonz\'{a}lez-Alfonso, E., Cernicharo, J., Alcolea, J., \& Orlandi, M.~A.} 1998, \textit{A\&A}, 334, 1016



\bibitem[Gray \etal\ 1995]{Gray1995}
{Gray, M.~D., Ivison, R.~J., Yates, J.~A., Humphreys, E.~M.~L.,
	Hall, P.~J., \& Field, D.} 1995, \textit{MNRAS}, 277, L67

\bibitem[Gray, Humphreys \& Yates 1999]{Gray1999}
{Gray, M.~D., Humphreys, E.~M.~L., \& Yates, J.~A} 1999, \textit{MNRAS}, 304, 906

\bibitem[Greenhill \etal\ 2007]{Greenhill2007}
{Greenhill, L.~J.,  \etal\ } 2007, in prep

\bibitem[Henkel, Matthews \& Morris 1983]{Henkel1983}
{Henkel, C., Matthews, H.~E., \& Morris, M.} 1983, \textit{ApJ}, 267, 184

\bibitem[Herpin \& Baudry 2000]{Herpin2000}
{Herpin, F. \& Baudry, A.} 2000, \textit{A\&A}, 359, 1117



\bibitem[Humphreys \etal\ 1997]{Humphreys1997}
{Humphreys, E.~M.~L., Gray, M.~D., Yates, J.~A., Field, D.} 1997, \textit{MNRAS}, 287, 663

\bibitem[Humphreys \etal\ 2001]{Humphreys2001}
{Humphreys, E.~M.~L., Yates, J.~A., Gray, M.~D., Field, D., Bowen, G.~H.} 2001, \textit{A\&A}, 379, 501

\bibitem[Humphreys \etal\ 2005]{Humphreys2005}
{Humphreys, E.~M.~L., Greenhill, L.~J., Reid, M.~J., Beuther, H., Moran, J.~M.,
 Gurwell, M., Wilner, D.~J., Kondratko, P.~T.} 2005, \textit{ApJ}, 634, L133

\bibitem[Ivison \etal\ 1998]{Ivison1998}
{Ivison, R.~J., Yates, J.~A., Hall, P.~J.} 1998, \textit{MNRAS}, 295, 813


\bibitem[Jewell \etal\ 1987]{Jewell1987}
{Jewell, P.~R. and Dickinson, D.~F. and Snyder, L.~E. and Clemens, D.~P.} 1987, \textit{ApJ}, 323, 749
 
\bibitem[Kalenskii, Slysh \& Val'Tts (2002)]{Kalenskii2002}
{Kalenskii, S.~V., Slysh, V.~I., Val'Tts, I.~E.} 2002, \textit{ARep}, 46, 49

\bibitem[Kemball \& Diamond 1997]{Kemball1997}
{Kemball, A. ~J., \& Diamond, P. ~J.} 1997, \textit{ApJ}, 481, L111

\bibitem[Kondratko, Greenhill \& Moran (2005)]{Kondratko2005}
{Kondratko, P.~T., Greenhill, L.~J., \& Moran, J.~M.} 2005, \textit{ApJ}, 618, 618

\bibitem[Krichbaum \etal\ (2007)]{Krichbaum2007}
{Krichbaum, T.~P., Graham, D.~A., Witzel, A., \etal\ } 2007, 
in: \textit{Towards the Event Horizon: High Resolution VLBI Imaging of Nuclei of Active Galaxies}, 
Exploring the Cosmic Frontier, ESO Astrophysics Symposia European Southern Observatory, Springer, 
p.\ 189

\bibitem[Martin-Pintado \etal\ 1989]{Martin-Pintado1989}
{Martin-Pintado, J., Bachiller, R., Thum, C., Walmsley, M.} 1989, 215, L13

\bibitem[Menten \& Melnick 1989]{Menten1989}
{Menten, K.~M. \& Melnick, G.~J.} 1989, \textit{ApJ}, 341, L91

\bibitem[Melnick \etal\ 1993]{Melnick1993}
{Melnick, G.~J., Menten, K.~M., Phillips, T.~G., Hunter, T.} 1993, \textit{ApJ}, 416, L37

\bibitem[Menten \etal\ 1990b]{Menten1990b}
{Menten, K.~M., Melnick, G.~J., Phillips, T.~G.} 1990b, \textit{ApJ}, 350, L41

\bibitem[Menten \etal\ 1990a]{Menten1990a}
{Menten, K.~M., Melnick, G.~J., Phillips, T.~G., Neufeld, D.~A.} 1990a, \textit{ApJ}, 363, L27

\bibitem[Menten \& Young (1995)]{Menten1995}{Menten, K.~M., \& Young, K.} 1995, \textit{ApJ}, 450, L67

\bibitem[Menten \etal\ 2006]{Menten2006}
{Menten, K.~M., Philipp, S.~D., G{\"u}sten, R., Alcolea, J., Polehampton, E.~T., Br{\"u}nken, S.} 2006, 
\textit{A\&A}, 454, L107


\bibitem[Neufeld \& Melnick 1991]{Neufeld1991}
{Neufeld, D.~A., Melnick, G.~J.} 1991, \textit{ApJ}, 368, 215



\bibitem[Pardo \etal\ 1998]{Pardo1998}
{Pardo, J.~R., Cernicharo, J., Gonzalez-Alfonso, E. and Bujarrabal, V.} 1998, A\&A, 329, 219



\bibitem[Patel \etal\ 2007]{Patel2007}
{Patel, N.~A., Curiel, S., Zhang, Q., Sridharan, T.~K.,
	Ho, P.~T.~P., \& Torrelles, J.~M.} 2007, \textit{ApJ}, 658, L55


\bibitem[Phillips, Kwan \& Huggins (1980)]{Phillips1980}
{Phillips, T.~G., Kwan, J., Huggins, P.~J.} 1980, 
in: B.~H.  Andrew (eds.),
\textit{Interstellar Molecules},
IAU Symposium, vol.\ 87, p.\ 21 


\bibitem[Planesas, Martin-Pintado \& Serabyn 1992]{Planesas1992}
Planesas, P., Martin-Pintado, J., Serabyn, E., 1992, \textit{ApJ}, 386, L23



\bibitem[Schilke \etal\ 2000]{Schilke2000}
{Schilke, P., Mehringer, D.~M., \& Menten, K.~M.} 2000, \textit{ApJ}, 528, L37


\bibitem[Schilke \etal\ 2001]{Schilke2001}
{Schilke, P., Benford, D.~J., Hunter, T.~R., Lis, D.~C., Phillips, T.~G.} 2001, \textit{ApJS}, 132, 281
 


\bibitem[Schilke \& Menten 2003]{Schilke2003}
{Schilke, P., \& Menten, K.~M.} 2003, \textit{ApJ}, 583, 446

\bibitem[Seaquist \etal\ 1996]{Seaquist1996}
{Seaquist, E.~R., Carlstrom, J.~E., Bryant, P.~M., Bell, M.~B.} 1996, \textit{ApJ}, 465, 691


\bibitem[Shinnaga \etal\ 2004]{Shinnaga2004}
{Shinnaga, H., Moran, J.~M., Young, K.~H., Ho, P.~T.~P.} 2004, \textit{ApJ}, 616, L47


\bibitem[Sobolev \etal\ (2002)]{Sobolev2002}
{Sobolev, A.~M., Ostrovskii, A.~B., Malyshev, A.~V.,
	Cragg, D.~M., Godfrey, P.~D., Sutton, E.~C., 
	Watson, W.~D., Ellingsen, S.~P., Caswell, J.~L.} 2002,
in: V. Migenes \& M.~J. Reid  (eds.),
\textit{Cosmic Masers: From Proto-Stars to Black Holes},
IAU Symposium, vol.\ 206, p.\ 179 


\bibitem[Spaans \& Norman 1997]{Spaans1997}
{Spaans, M., \& Norman, C.~A.} 1997, \textit{ApJ}, 488, 27 

\bibitem[Strelnitski \etal\ 1996]{Strelnitski1996}
{Strelnitski, V.~S., Ponomarev, V.~O., Smith, H.~A.} 1996, \textit{ApJ}, 470, 1118

\bibitem[Tennyson \etal\ 2001]{Tennyson2001}
{Tennyson, J., Zobov, N. F., Williamson, R., Polyansky, O.L., Bernath, P. F.} 2001,
\textit{J. Phys. Chem. Ref. Data}, 30, 735

\bibitem[Thum \etal\ 1994a]{Thum1994a}
{Thum, C., Matthews, H.~E., \& Martin-Pintado, J., 
Serabyn, E., Planesas, P., Bachiller, R.} 1994a, \textit{A\&A}, 283, 582

\bibitem[Thum \etal\ 1994b]{Thum1994b}
{Thum, C., Matthews, H.~E., Harris, A.~I., Tacconi, L.~J., Schuster, K.~F., Martin-Pintado, J.} 1994b, 
\textit{A\&A}, 288, L25

\bibitem[Thum \etal\ 1995]{Thum1995}
{Thum, C., Strelnitski, V.~S., Martin-Pintado, J., Matthews, H.~E., Smith, H.~A.} 1995, \textit{A\&A}, 300, 843

\bibitem[Thum \& Morris 1999]{Thum1999}
{Thum, C. \& Morris, D.} 1999, \textit{A\&A}, 344, 923

\bibitem[Waters \etal\ 1980]{Waters1980}
{Waters, J.~W., Kakar, R.~K., Kuiper, T.~B.~H.,
	Roscoe, H.~K., Swanson, P.~N., Rodriguez Kuiper, E.~N.,
	Kerr, A.~R., Thaddeus, P., Gustincic, J.~J.} 1980, \textit{ApJ}, 235, 57

\bibitem[Weintroub \etal\ 2007]{Weintroub2007}
{Weintroub, J., \etal\  } 2007, in prep

\bibitem[Yates \etal\ 1995]{Yates1995}
{Yates, J.~A., Cohen, R.~J., Hills, R.~E.}, 1995, \textit{MNRAS}, 273, 529

\bibitem[Yates \etal\ 1996]{Yates1996}
{Yates, J.~A. \& Cohen, R.~J.} 1996, \textit{MNRAS}, 278, 655

\bibitem[Yates \etal\ 1997]{Yates1997}
 {Yates, J.A., Field, D., \& Gray, M.D.} 1997, \textit{MNRAS}, 285, 303


\end{thebibliography}
\end{document}